\documentstyle[prl,aps]{revtex}
\begin{document}
\draft
\title{Generalizing $\Phi-$measure of event-by-event fluctuations \\
in high-energy heavy-ion collisions
}
\author{Stanis\l aw Mr\' owczy\' nski\footnote{Electronic address:
{\tt mrow@fuw.edu.pl}}}

\address{So\l tan Institute for Nuclear Studies, \\
ul. Ho\.za 69, PL - 00-681 Warsaw, Poland \\
and Institute of Physics, Pedagogical University, \\
ul. Konopnickiej 15, PL - 25-406 Kielce, Poland}

\date{10-th May 1999, revised 18-th July 1999}

\maketitle

\begin{abstract}

The $\Phi-$measure of event-by-event fluctuations in high-energy 
heavy-ion collisions corresponds to the second moment of the 
fluctuating quantity distribution of interest. It is shown that the measure 
based on the third moment preserves the properties of $\Phi$ but those 
related to the higher moments do not. In particular, only the second and 
third moment measures are intensive as thermodynamic quantities. The 
$\Phi_2-$ and $\Phi_3-$measure of  $p_{\perp}-$fluctuations are computed 
for the hadron gas in equilibrium and the results are analyzed in context of 
the experimental data.

\end{abstract}

\vspace{0.5cm}
 

\vspace{0.5cm}

Large acceptance detectors allow one for a detailed analysis of 
individual collisions of heavy-ions at high-energies. Due to hundreds
or even thousands of particles produced in these collisions, variety 
of statistical methods can be applied. There are several interesting 
proposals \cite{Gaz92,Sto95,Shu98,Mro98a,Ste99} to use the fluctuation 
measurements as a potential source of information on the collision 
dynamics. However, one faces a problem how to disentangle the 
`dynamical' fluctuations from the `trivial' geometrical ones due to the 
impact parameter variation. The latter fluctuations are large and dominate 
the fluctuations of  all extensive event characteristics such as multiplicity 
or transverse energy. Using the fluctuation (or correlation) measure $\Phi$, 
which was introduced in our paper \cite{Gaz92}, resolves the problem 
in a specific way. By construction, $\Phi$ is exactly the same for 
nucleon-nucleon (N--N) and nucelus-nucelus (A--A) collisions if 
the A--A collision is a simple superposition of N--N interactions. 
Consequently, $\Phi$ is independent of the centrality of A--A collision
in such a case. On the other hand, $\Phi$ equals zero when the inter-particle 
correlations are entirely absent. The $\Phi-$measure can be applied to the 
fluctuations of  kinematical quantities such as the event energy or 
transverse momentum and to the fluctuations of event chemical 
composition \cite{Gaz99b,Mro99}. 

The NA49 Collaboration plans to study the chemical fluctuations 
in a near future \cite{NA498} while the data on the transverse momentum 
fluctuations have been already published \cite{Rol98,App99}. The value 
of $\Phi_{p_{\perp}}$ in the central Pb--Pb collisions at 158 GeV per 
nucleon has appeared to be smaller than expected. It has been also claimed 
\cite{App99} that the correlations, which are of short range in the momentum 
space, are responsible for the nonzero positive value of  $\Phi_{p_{\perp}}$ 
being observed.  The result has been widely discussed 
\cite{Ble98,Liu99,Gaz99a,Mro98b,Alb99,Cap99}. In particular, our calculations 
of $\Phi_{p_{\perp}}$ in the equilibrium hadron gas show \cite{Mro98b} that the 
positive value of $\Phi_{p_{\perp}}$ appears due to the boson statistics of pions. 
When the hadronic system at freeze-out is identified with the pion gas, 
the calculated $\Phi_{p_{\perp}}$ slightly overestimates the experimental value 
\cite{App99} but, as discussed here, the inclusion of the pions which come 
from the resonance decays removes the discrepency.

The $\Phi-$measure corresponds to the second moment of the fluctuating
quantity, say the event transverse momentum. Recently, it has been
suggested \cite{Bel99} to use the higher moments in an analogous way.
However, the authors of \cite{Bel99} have not realized that the fluctuation 
measures based on the higher moments, except that of the third one, do not 
posses a key property of $\Phi$ which has been mentioned above. Namely, 
$\Phi_{NN} = \Phi_{AA}$ if the A--A collision is a simple superposition 
of N--N interactions. When treated as thermodynamic quantities, the second
and third moment measures are intensive while the higher moment ones are
not. The aim of this note is to substantiate the comment and to discuss 
usefulness of the third moment measure. We focus our attention on the 
$p_{\perp}-$fluctuations which have been already studied experimentally \cite{Rol98,App99}.

Let us first introduce the $\Phi-$measure. One defines the single-particle 
variable $z \buildrel \rm def \over = x - \overline{x}$ with the overline 
denoting averaging over a single particle inclusive distribution. The event 
variable $Z$, which is a multiparticle analog of $z$, is defined as 
$Z \buildrel \rm def \over = \sum_{i=1}^{N}(x_i - \overline{x})$, 
where the summation runs over particles from a given event.
By construction, $\langle Z \rangle = 0$, where $\langle ... \rangle$ 
represents averaging over events. Finally, the $\Phi-$measure is defined 
in the following way
\begin{equation}\label{phi}
\Phi \buildrel \rm def \over = 
\sqrt{\langle Z^2 \rangle \over \langle N \rangle} -
\sqrt{\overline{z^2}} \;.
\end{equation}
There is an obvious generalization of the definition (\ref{phi}) suggested 
in \cite{Bel99}. Namely,
\begin{equation}\label{phi-n}
\Phi_n \buildrel \rm def \over = 
\bigg( { \langle Z^n \rangle \over \langle N \rangle }\bigg)^{1/n} -
\big( \overline{z^n}\big)^{1/n} \;.
\end{equation}

The fact that $\Phi_n = 0$, when no inter-particle correlations are 
present, is evident \cite{Gaz92,Bel99}.  We are now going to show that 
$\Phi_3$ as $\Phi_2$, in contrast to $\Phi_n$ with $n >3$, possesses 
another nontrivial property which is so useful in the data analysis. Namely, 
$\Phi_2$ and $\Phi_3$ are {\it independent} of the source number distribution 
if the particles originate from several identical sources. Then, $\Phi_2$ and
$\Phi_3$ are independent of the impact parameter if the A--A collision is 
a superposition of N--N interactions. Let us prove this property.

$P_1(X)$ is the normalized distribution of 
$X \buildrel \rm def \over = \sum_{i=1}^{N}x_i$, when the particles 
come from the single source. If we have $k$ sources distributed
according to $p_k$, the $X-$distribution reads
$$
P(X) = \sum_{k=1}^{\infty}p_k \int dX_1\, .....\; dX_k \; 
P_1(X_1) \,.....\,P_1(X_k) \; \delta(X - (X_1 + ..... + X_k)) \;.
$$
The moments of $P(X)$ are
\begin{equation}\label{moments}
\langle X^n \rangle \buildrel \rm def \over = \int dX \, X^n P(X) = 
(-i)^n{d^n \over dQ^n} {\cal F}(Q)\Bigg|_{Q=0} \;,
\end{equation}
where the generating function ${\cal F}$ equals
$$
{\cal F}(Q) \buildrel \rm def \over = \int dX \; e^{iQX} P(X) =
\sum_{k=1}^{\infty}p_k \Big[{\cal F}_1(Q)\Big]^k \;
$$
with 
$$
{\cal F}_1(Q) \buildrel \rm def \over = \int dX \; e^{iQX} P_1(X) \;.
$$
Using eq.~(\ref{moments}), one computes the first five moments of
$P(X)$ as
\begin{eqnarray*}
\langle X \rangle &=& \langle k \rangle \langle X \rangle_1 \;,
\\[2mm]
\langle X^2 \rangle &=& \langle k \rangle \langle X^2 \rangle_1 
+\langle k(k-1) \rangle \langle X \rangle_1^2 \;,
\\[2mm]
\langle X^3 \rangle &=& \langle k \rangle \langle X^3 \rangle_1 
+3\langle k(k-1) \rangle \langle X^2 \rangle_1 \langle X \rangle_1 
+\langle k(k-1)(k-2) \rangle \langle X \rangle_1^3 \;,
\\[2mm]
\langle X^4 \rangle &=& \langle k \rangle \langle X^4 \rangle_1 
+4\langle k(k-1) \rangle \langle X^3 \rangle_1 \langle X \rangle_1 
+3\langle k(k-1) \rangle \langle X^2 \rangle_1^2 
\\&&
+3\langle k(k-1)(k-2) \rangle \langle X^2 \rangle_1 
\langle X \rangle_1^2 
+\langle k(k-1)(k-2)(k-3) \rangle \langle X \rangle_1^4 \;,
\\[2mm]
\langle X^5 \rangle &=& \langle k \rangle \langle X^5 \rangle_1 
+5\langle k(k-1) \rangle \langle X^4 \rangle_1 \langle X \rangle_1 
+10\langle k(k-1) \rangle \langle X^3 \rangle_1\langle X^2 \rangle_1 
\\&&
+7\langle k(k-1)(k-2)\rangle \langle X^3 \rangle_1\langle X \rangle_1^2
+9\langle k(k-1)(k-2)\rangle \langle X^2 \rangle_1^2\langle X \rangle_1
\\&&
+7\langle k(k-1)(k-2)(k-3)\rangle \langle X^2 \rangle_1\langle X \rangle_1^3
+\langle k(k-1)(k-2)(k-3)(k-4)\rangle \langle X \rangle_1^5 \;,
\end{eqnarray*}
where
$$
\langle X^n \rangle_1 \buildrel \rm def \over = \int dX \, X^n P_1(X) 
\;\;\;\;\;\;\;\;\;\;\;\; {\rm and} \;\;\;\;\;\;\;\;\;\;\;\;
\langle k^n \rangle \buildrel \rm def \over = \sum_{k=1}^{\infty}k^n p_k \;.
$$
Applying these formulas to the variable $Z$ and taking into account that
by definition $\langle Z \rangle = \langle Z \rangle_1 = 0$, we get
\begin{eqnarray*}
\langle Z^2 \rangle &=& \langle k \rangle \langle Z^2 \rangle_1 \;, 
\\[2mm]
\langle Z^3 \rangle &=& \langle k \rangle \langle Z^3 \rangle_1  \;,
\\[2mm]
\langle Z^4 \rangle &=& \langle k \rangle \langle Z^4 \rangle_1 
+3\langle k(k-1) \rangle \langle Z^2 \rangle_1^2 \;,
\\[2mm]
\langle Z^5 \rangle &=& \langle k \rangle \langle Z^5 \rangle_1 
+10\langle k(k-1) \rangle \langle Z^3 \rangle_1\langle Z^2 \rangle_1 \;.
\end{eqnarray*}
Since $\langle N \rangle = \langle k \rangle \langle N \rangle_1$, one
finds that
\begin{eqnarray*}
{\langle Z^2 \rangle \over \langle N \rangle}
= {\langle Z^2 \rangle_1 \over \langle N \rangle_1} 
\;, \;\;\;\;\;\;\;\;\;\;\;\;\;
{\langle Z^3 \rangle \over \langle N \rangle}
= {\langle Z^3 \rangle_1 \over \langle N \rangle_1} \;, 
\end{eqnarray*}
but analogous formulas do not hold for $\langle Z^4\rangle$ and 
$\langle Z^5 \rangle$. Instead,
\begin{eqnarray*}
{\langle Z^4 \rangle \over \langle N \rangle}
&=& {\langle Z^4 \rangle_1 \over \langle N \rangle_1}
+3{\langle k(k-1) \rangle \over \langle k \rangle}
{\langle Z^2 \rangle_1^2 \over \langle N \rangle_1} \;,
\\[2mm]
{\langle Z^5 \rangle \over \langle N \rangle}
&=& {\langle Z^5 \rangle_1 \over \langle N \rangle_1} 
+10{ \langle k(k-1)\rangle \over \langle k \rangle}
{\langle Z^3 \rangle_1\langle Z^2 \rangle_1 \over \langle N \rangle_1} \;.
\end{eqnarray*}
Therefore, $\Phi_2$ and $\Phi_3$ are independent of the source number
distribution while $\Phi_4, \;\; \Phi_5$ and, obviously, $\Phi_n$ with 
$n> 5$ do depend on $p_k$. The inclusive distribution, which determines
$\overline{z^n}$, is, of course, independent of the source distribution.
The above results also show that $\Phi_2$ and $\Phi_3$ are intensive 
quantities, i.e. they are independent of the system size, while $\Phi_n$ 
with $n > 3$ are not. Indeed, when the source number is fixed,  
$\langle k^l \rangle =k^l $ and one observes that only $\Phi_2$ and 
$\Phi_3$ do not dependent on $k$. Let us note here that the 
independence of $k$ is, in principle, a weaker requirement than the
independence of $p_k$.

The $\Phi_2-$measure is sensitive to the fluctuations or correlations of 
various origin. For example, it acquires a finite value, which is positive 
for bosons and negative for fermions, due to the quantum statistics 
\cite{Mro98b}. The correlation between the particle multiplicity and 
their kinematical characteristics also influences $\Phi_2$ \cite{Gaz92}. 
The energy-momentum conservation and presence of the collective 
motion introduces additional inter-particle correlations. Thus, one 
concludes that the nonvanishing value of $\Phi_2$ signals the existence 
of the correlations in the system but it does not explain their origin. 
In such a situation, $\Phi_3$ seems to be very useful. Indeed, simultaneous 
measurements of $\Phi_2$ and $\Phi_3$ might help to identify  the 
fluctuations which dominate in the system. For this purpose one should 
theoretically estimate contributions of various correlations to $\Phi_2$ 
and $\Phi_3$ 

In our paper \cite{Mro98b} we have discussed how to compute $\Phi_2$ 
in the ideal quantum gas. Now, we are going to extend these calculations 
to the case of $\Phi_3$. For comparison, we also present here the earlier 
published \cite{Mro98b} results on $\Phi_2$. At first, the energy fluctuations 
are considered. Therefore, the single particle variable $x$ is identified with 
the particle energy $E$. Then, one immediately finds that
\begin{equation}\label{phiE1}
\overline{z^n} = {1 \over \rho}\int{d^3p \over (2\pi )^3}
\,(E - \overline{E})^n{1\over \lambda^{-1}e^{\beta E} \pm 1} \;,
\end{equation}
where the single particle average energy is
$$
\overline E = {1 \over \rho}\int{d^3p \over (2\pi )^3} \;
{E \over \lambda^{-1}e^{\beta E} \pm 1} \;,
$$
while the particle density $\rho$ equals
\begin{equation}\label{rho}
\rho = \int{d^3p \over (2\pi )^3} \;
{1 \over \lambda^{-1}e^{\beta E} \pm 1} \;;
\end{equation}
$\beta \equiv T^{-1}$ is the inverse temperature; 
$\lambda \equiv e^{\beta \mu}$ denotes the fugacity and $\mu$ the chemical 
potential; $E \equiv \sqrt{m^2 + {\bf p}^2}$ with $m$ being the particle
mass and ${\bf p}$ its momentum; the upper sign is for fermions while the
lower one for bosons. 

Since $Z = U - N\overline{E}$, where $U$ is the system energy,
$\langle Z^2 \rangle$ and $\langle Z^3 \rangle$ are computed as
\begin{eqnarray}\label{Z2}
\langle Z^2 \rangle &=& {1 \over \Xi}
\Bigg[ {\partial^2 \over  \partial \beta^2}
+2 \overline{E} \, \lambda {\partial^2 \over \partial \beta \,\partial\lambda}
+\overline{E}^2 \bigg(\lambda {\partial \over \partial \lambda}\bigg)^2 \Bigg]
\,\Xi(V,T,\lambda) \;,
\\[3mm]
\langle Z^3 \rangle &=& -{1 \over \Xi}
\Bigg[ {\partial^3 \over  \partial \beta^3}
+3 \overline{E} \,{\partial^2 \over\partial \beta^2}
\lambda {\partial \over\partial\lambda}
+3 \overline{E}^2\,{\partial \over\partial \beta}
\bigg(\lambda {\partial \over\partial\lambda}\bigg)^2
+\overline{E}^3\bigg(\lambda {\partial \over \partial \lambda}\bigg)^3 \Bigg]
\,\Xi(V,T,\lambda) \;, \nonumber
\end{eqnarray}
where $\Xi(V,T,\lambda)$ is the grand canonical partition function
\cite{Hua63} defined as
$$
\Xi(V,T,\lambda) = \sum_N \sum_{\alpha} 
\lambda^N e^{-\beta U_{\alpha}} \;,
$$
with $V$ denoting the system volume and the index $\alpha$ numerating 
the system quantum states\footnote{The formulas from \cite{Mro98b} 
analogous to (\ref{Z2}) and (\ref{Xi}) are erroneously written but the 
final results are correct.}. As well known \cite{Hua63}, the grand 
canonical partition function of the quantum ideal gas equals
\begin{equation}\label{Xi}
{\rm ln}\,\Xi(V,T,\lambda) = \pm g\,V  \int{d^3p \over (2\pi )^3} \;
{\rm ln}\big[1 \pm \lambda\,e^{-\beta E} \big] \;,
\end{equation}
with $g$ being the number of the particle internal degrees of freedom.
After a rather lengthy calculation, one finds
\begin{equation}\label{phiE2}
{\langle Z^2 \rangle \over \langle N \rangle }= 
{1 \over \rho}\int{d^3p \over (2\pi )^3}
\,(E - \overline{E})^2 \; {\lambda^{-1}e^{\beta E}
\over (\lambda^{-1}e^{\beta E} \pm 1)^2} \;,
\end{equation}
and 
\begin{equation}\label{phiE3}
{\langle Z^3 \rangle \over \langle N \rangle }= 
{1 \over \rho}\int{d^3p \over (2\pi )^3}
\,(E - \overline{E})^3 \;
{\lambda^{-1}e^{\beta E} (\lambda^{-1}e^{\beta E}\mp 1)
\over (\lambda^{-1}e^{\beta E} \pm 1)^3} \;.
\end{equation}
As expected, $\Phi_2$ and $\Phi_3$, which are given 
by the formulas (\ref{phiE1},\ref{phiE2},\ref{phiE3}), are
intensive thermodynamic quantities, i.e. they are independent 
of the system volume. We also note that $\Phi_2$ and $\Phi_3$
are independent of $g$. One observes that the sign of $\Phi_2$ is 
definite i.e. $\Phi_2 < 0$ for fermions, $\Phi_2 > 0$ for bosons 
and $\Phi_2 = 0$ in the classical limit ($\lambda^{-1} \gg 1$)
\cite{Mro98b}. The sign of $\Phi_3$ is not definite but $\Phi_3$ 
still vanishes for the classical particles. 

When the particles are massless and their chemical potential vanish
($\lambda = 1$), the calculations can be performed analytically to
the end. Then, eqs. (\ref{phiE1},\ref{phiE2},\ref{phiE3}) give
$$
\Phi_2 \cong {-0.07 \choose \;\;\;0.40} \,T \;,
\;\;\;\;\;\;\;\;
\Phi_3 \cong {0.11 \choose -\infty} \,T \;,
$$
where the upper case is for fermions and the lower one for bosons.
For $m = \mu = 0$ the bosonic $\Phi_3$ appears to be (logarithmically) 
divergent due to the singular character of the function 
$(e^{\beta E} - 1)^{-3}$ at $E \rightarrow 0$.

One immediately modifies eqs.~(\ref{phiE1},\ref{phiE2},\ref{phiE3}) 
for the case of the transverse momentum. The respective equations read:
\begin{equation}\label{phipT1}
\overline{z^n} = {1 \over \rho}\int{d^3p \over (2\pi )^3} \,
(p_{\perp} - \overline{p}_{\perp})^n 
{1\over \lambda^{-1}e^{\beta E} \pm 1} \;,
\end{equation}
\begin{equation}\label{phipT2}
{\langle Z^2 \rangle \over \langle N \rangle }= 
{1 \over \rho}\int{d^3p \over (2\pi )^3}
\,(p_{\perp} - \overline{p}_{\perp})^2 \; {\lambda^{-1}e^{\beta E}
\over (\lambda^{-1}e^{\beta E} \pm 1)^2} \;,
\end{equation}
\begin{equation}\label{phipT3}
{\langle Z^3 \rangle \over \langle N \rangle }= 
{1 \over \rho}\int{d^3p \over (2\pi )^3}
\,(p_{\perp} - \overline{p}_{\perp})^3 \;
{\lambda^{-1}e^{\beta E} (\lambda^{-1}e^{\beta E}\mp 1)
\over (\lambda^{-1}e^{\beta E} \pm 1)^3} \;,
\end{equation}
where $p_{\perp} = p \, {\rm sin}\Theta$ with $p \equiv \vert {\bf p} \vert$ 
and $\Theta$ being the angle between ${\bf p}$ and the beam $(z)$ axis, and 
$$
\overline{p}_{\perp} = {1 \over \rho}\int{d^3p \over (2\pi )^3} \;
{p_{\perp} \over \lambda^{-1}e^{\beta E} \pm 1} \;.
$$

In Figs. 1 and 2 we present with dashed lines the $\Phi_2-$ and 
$\Phi_3-$measure of $p_{\perp}-$fluctuations in the ideal pion gas. 
The pions are, of course, massive ($m_{\pi}=140$ MeV), 
so $\Phi_2$ and $\Phi_3$ are found numerically from 
eqs.~(\ref{phipT1},\ref{phipT2},\ref{phipT3}). The calculations are 
performed for several values of the pion chemical potential. In the chemical
equilibrium $\mu = 0$. As seen, $\Phi_2$ is positive but $\Phi_3$ is negative. 
At $T \cong 200$ MeV and $\mu = 70$ MeV, $\Phi_3$ experiences a rapid
growth. This happens because the first term from eq.~(\ref{phi-n}) 
changes the sign from positive to negative at $T \cong 200$ MeV. 

It is a far going idealization to treat a fireball at freeze-out 
as an ideal gas of pions. A substantial fraction of the final state 
pions come from the hadron resonances. These pions do not `feel' the 
Bose-Einstein statistics at freeze-out and consequently the values
of $\Phi_2$ and $\Phi_3$ should be significantly reduced. We estimate 
the role of resonances in the following way. The spectrum of pions, 
which originate from the resonance decays, is not dramatically different 
than that given by the equilibrium distribution \cite{Sol91}. Therefore, 
we treat the fireball at freeze-out as a mixture of `quantum' pions - those 
called `direct' - and the `classical'  pions which come from the resonance 
decays. Since the weighting functions in 
eqs. (\ref{phipT1},\ref{phipT2},\ref{phipT3}) are all equal to 
$\lambda e^{-\beta E}$ in the classical limit, the formulas 
analogous to (\ref{phipT1},\ref{phipT2},\ref{phipT3}) are
\begin{equation}\label{phipTr1}
\overline{z^n} = {1 \over \rho}\int{d^3p \over (2\pi )^3} \,
(p_{\perp} - \overline{p}_{\perp})^n 
\bigg[{1\over \lambda^{-1}e^{\beta E} - 1} 
+ \lambda_r e^{-\beta E} \bigg] \;,
\end{equation}
\begin{equation}\label{phipTr2}
{\langle Z^2 \rangle \over \langle N \rangle }= 
{1 \over \rho}\int{d^3p \over (2\pi )^3}
\,(p_{\perp} - \overline{p}_{\perp})^2 \; 
\bigg[{\lambda^{-1}e^{\beta E}
\over (\lambda^{-1}e^{\beta E} - 1)^2}
+ \lambda_r e^{-\beta E} \bigg] \;,
\end{equation}
\begin{equation}\label{phipTr3}
{\langle Z^3 \rangle \over \langle N \rangle }= 
{1 \over \rho}\int{d^3p \over (2\pi )^3}
\,(p_{\perp} - \overline{p}_{\perp})^3 \;
\bigg[ {\lambda^{-1}e^{\beta E} (\lambda^{-1}e^{\beta E}+ 1)
\over (\lambda^{-1}e^{\beta E} - 1)^3} 
+ \lambda_r e^{-\beta E} \bigg] \;,
\end{equation}
with
$$
\overline{p}_{\perp} = {1 \over \rho}\int{d^3p \over (2\pi )^3} \;
p_{\perp}\bigg[{1 \over \lambda^{-1}e^{\beta E} - 1} 
+ \lambda_r e^{-\beta E} \bigg] \;,
$$
$$
\rho = \int{d^3p \over (2\pi )^3} \;
\bigg[{1 \over \lambda^{-1}e^{\beta E} - 1} 
+ \lambda_r e^{-\beta E} \bigg] \;.
$$
The parameter $\lambda_r$ is chosen is such a way that the number of 
`classical' pions equals to the number of pions from the resonance
decays. Thus, $\lambda_r$ is temperature dependent. In the actual 
calculations, we have taken into account the lightest resonances:  
$\rho(770)$ and $\omega(782)$ which give the dominant contribution.
The life time of $\rho$, which is 1.3 fm/$c$, is not much longer
than the time of the fireball decoupling and some pions from 
the $\rho$ decays can still `feel' the effect of Bose statitistics. Therefore, 
the contribution of $\rho$ to the `classical' pions is presumably
overestimated in our calculations. Since we neglect the heavier
resonances  and weakly decaying particles, which also contribute
to the final state pions,  the two effects partially compensate each
other.  In any case, our calculations show that the resonances do
not change the values of $\Phi_2$ and $\Phi_3$ dramatically in
the domain of temperatures of interest. 

In Figs. 1 and 2 the solid lines represents $\Phi_2-$ and $\Phi_3-$measure 
which include the resonances. The chemical potentials of $\rho$ 
and $\omega$ are assumed to be equal to that of pions. As seen, the
role of the resonaces is negligible at the temperatures below 100 MeV
but above this temperature the resonances reduce the fluctuations noticeably.
As already mentioned, $\Phi_2-$measure of $p_{\perp}-$fluctuations has 
been experimentally measured in the central Pb--Pb collisions by the 
NA49 collaboration. The first result has been published as 
$\Phi_2= 0.7 \pm 0.5 $ MeV \cite{Rol98} but the value of  $\Phi_2$ is 
increased to $4.6 \pm 1.5 $ MeV when the two-track resolution effect is properly
taken into account \cite{App99}. If we identify the system freeze-out 
temperature with the slope parameter deduced from the pion transverse 
momentum distribution $T \cong 180$ MeV \cite{App98}. Then, the value 
of $\Phi_2$, which is read out from Fig.~1 for $\mu=0$, equals 15 MeV 
for no resonances and $\Phi_2= 8.7$ MeV when the resonances are included. 
The temperature is significantly reduced if the transverse hydrodynamic 
expansion is taken into account. The freeze-out temperature obtained by 
means of the simultaneous analysis of the single particle spectra and 
the two-particle correlations is about 120 MeV  \cite{App98}. Then, the 
value of  $\Phi_2$ for $\mu=0$ equals 6.5 MeV for the case of no resonances 
and $\Phi_2= 5.6$ MeV when the resonances are included. The latter number 
agrees perfectly well with the mentioned above experimental value. This 
strongly supports the claim \cite{App99} that the short range correlations 
due to the Bose-Einstein statistics of pions play a dominant role in the 
hadronic system produced in central heavy-ion collisions. However, it would 
be very interesting to check whether the experiment also confirms our 
prediction on $\Phi_3$ which is presented in Fig.~2. As seen, 
$\Phi_3 =-12.3$ MeV for $T=120$ MeV and $\mu= 0$ when the resonances are 
taken into account.

Let us close this paper with a technical remark. When the $\Phi-$measure 
is applied to the real data or simulated events, it is rather inconvenient to 
use the formula (\ref{phi}) because then one has to process the data twice; 
in the first run one evaluates the inclusive average $\overline{x}$ and then 
computes the moments of $Z$ and $z$. To avoid the double data processing 
one can use the formula derived in \cite{Liu99} which is
\begin{equation}\label{phi2}
\Phi_2 = \bigg( {\langle X^2 \rangle \over \langle N \rangle}
-{2\langle X \rangle \langle XN \rangle \over \langle N \rangle^2}
+{\langle X \rangle^2 \langle N^2 \rangle \over \langle N \rangle^3}
\bigg)^{1/2} -
\bigg( {\langle X_2 \rangle \over \langle N \rangle}
-{\langle X \rangle^2 \over \langle N \rangle^2} \bigg)^{1/2} \;,
\end{equation}
where the event variable $X_2$ is defined as 
$X_2 \buildrel \rm def \over = \sum_{i=1}^{N}x_i^2$. The expression of 
$\Phi_3$, which is analogous to (\ref{phi2}), reads
\begin{eqnarray*}
\Phi_3 = \bigg( {\langle X^3 \rangle \over \langle N \rangle}
-{3\langle X \rangle \langle X^2 N \rangle \over \langle N \rangle^2}
+{3\langle X \rangle^2 \langle XN^2 \rangle \over \langle N \rangle^3}
-{\langle X \rangle^3 \langle N^3 \rangle \over \langle N \rangle^4}
\bigg)^{1/3} -
\bigg( {\langle X_3 \rangle \over \langle N \rangle}
-{3\langle X_2 \rangle \langle X \rangle \over \langle N \rangle^2}
+{2\langle X \rangle^3 \over \langle N \rangle^3} \bigg)^{1/3} \;,
\end{eqnarray*}
with $X_3 \buildrel \rm def \over = \sum_{i=1}^{N}x_i^3$.

We conclude our study as follows. The $\Phi_3-$measure, which is based
on the third moment of the fluctuating quantity distribution, preserves the 
advantageous properties of  $\Phi_2$ while the higher moment measures do
not. Simultaneous usage of $\Phi_2$ and $\Phi_3$ may help in identifying 
the origin of correlations observed in the final state of heavy-ion collisions
at high-energies. In particular, the measurement of $\Phi_3$ of
$p_{\perp}-$fluctuations can decisively confirm that the dominant correlations 
in the central collisions are those of the quantum statistics.

\vspace{1cm}

I am very grateful to Marek Ga\' zdzicki for fruitful discussions
and critical reading of the manuscript.


\newpage
\vspace{1cm}
\begin{center}
{\bf Figure Captions}
\end{center}
\vspace{0.5cm}

\noindent
{\bf Fig. 1.} 
$\Phi_2-$measure of $p_{\perp}-$fluctuations in the hadron gas as a function 
of temperature for four values of the chemical potential. The resonances are 
either neglected (dashed lines) or taken into account (solid lines). The most 
upper dashed and solid lines correspond to $\mu = 70$ MeV, the lower ones 
to $\mu = 0$, etc. 

\vspace{0.5cm}

\noindent
{\bf Fig. 2.} 
$\Phi_3-$measure of $p_{\perp}-$fluctuations in the hadron gas as a function 
of temperature for four values of the chemical potential. The resonances are 
either neglected (dashed lines) or taken into account (solid lines). The most 
upper dashed and solid lines correspond to $\mu = 70$ MeV, the lower ones 
to $\mu = 0$, etc. 

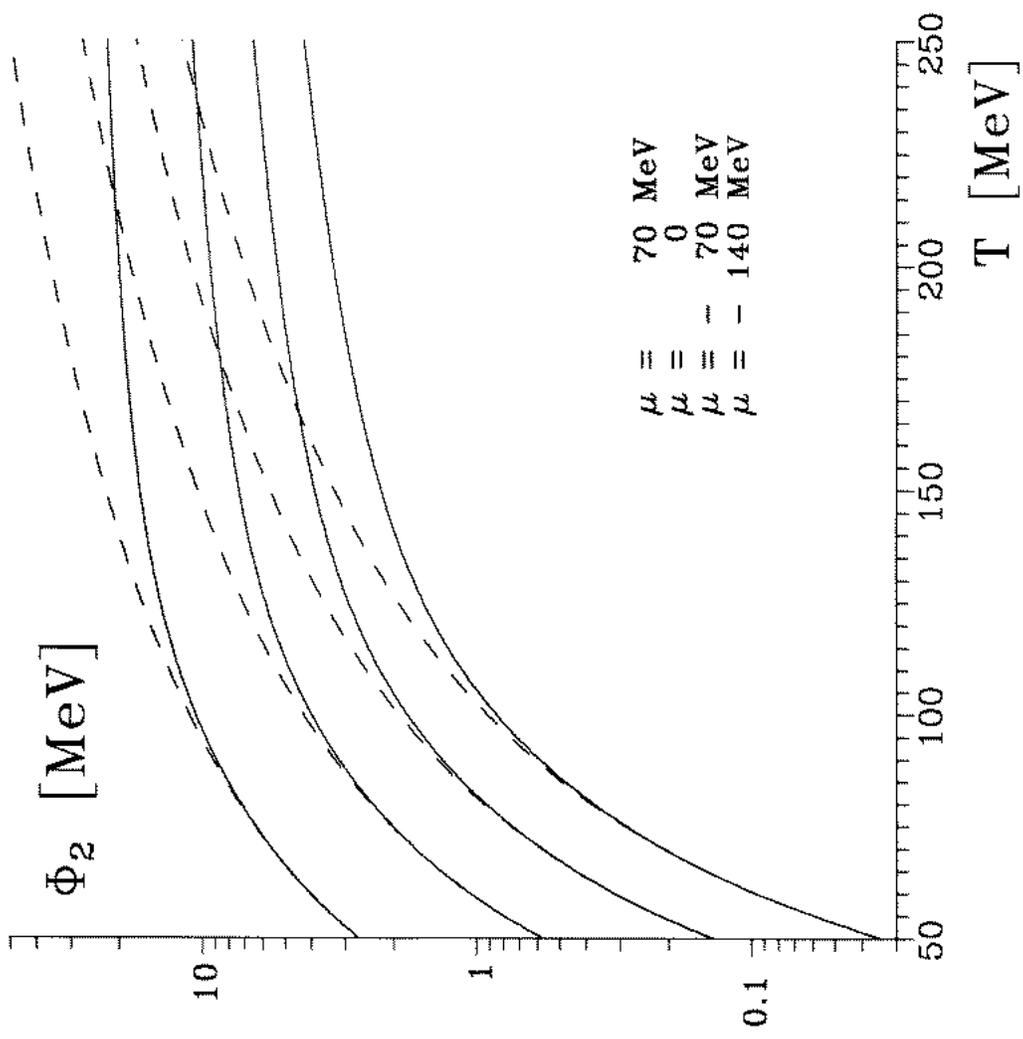
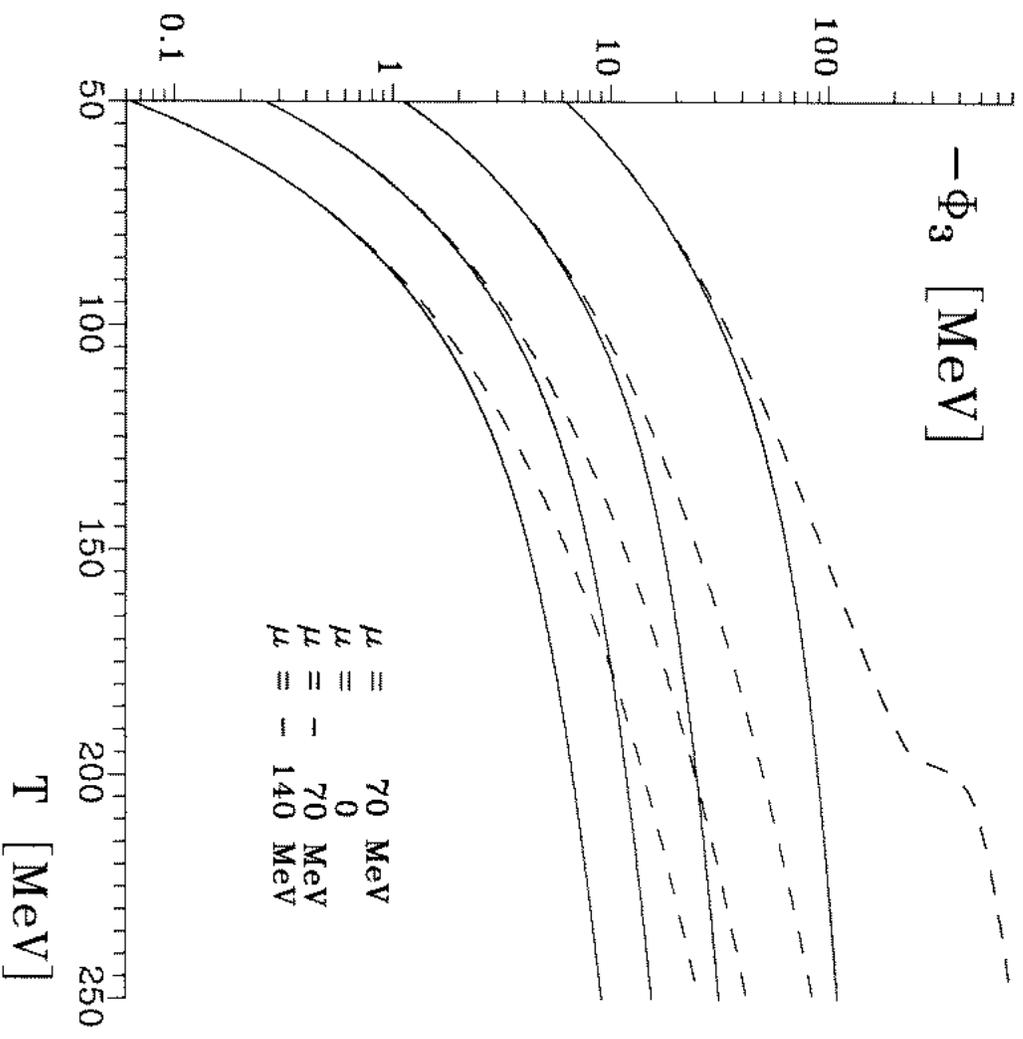
\end{document}